\documentclass[12pt,preprint]{aastex}
\usepackage[section] {placeins}
\usepackage{multirow}
\usepackage{amsmath}
\shorttitle{CONTINUUM OF ORION-KL}
\shortauthors{Friedel \& Widicus Weaver}
\begin{document}
\newcommand\dme{(CH$_3$)$_2$O}
\newcommand\mef{HCOOCH$_3$}
\newcommand\kms{km s$^{-1}$}
\newcommand\jbm{Jy bm$^{-1}$}
\newcommand\etcn{C$_2$H$_5$CN}
\newcommand\ace{(CH$_3$)$_2$CO}

\title{A HIGH SPATIAL RESOLUTION STUDY OF THE $\lambda$=3 MM CONTINUUM OF ORION-KL}

\author{D. N. Friedel\altaffilmark{1} and S. L. Widicus Weaver\altaffilmark{2}}

\altaffiltext{1}{Department of Astronomy, 1002 W. Green St., University of
Illinois, Urbana IL 61801\\
email: friedel@astro.illinois.edu}
\altaffiltext{2}{Department of Chemistry, Emory University, Atlanta, GA  30322\\
email: susanna.widicus.weaver@emory.edu}

\begin{abstract}
Recent interferometric observations have called into question the traditional view of the Orion-KL region, which displays one of the most well-defined cases of chemical differentiation in a star-forming region. Previous, lower-resolution images of Orion-KL show emission signatures for oxygen-bearing organic molecules toward the Orion Compact Ridge, and emission for nitrogen-bearing organic molecules toward the Orion Hot Core.  However, more recent observations at higher spatial resolution indicate that the bulk of the molecular emission is arising from many smaller, compact clumps that are spatially distinct from the traditional Hot Core and Compact Ridge sources. It is this type of observational information that is critical for guiding astrochemical models, as the spatial distribution of molecules and their relation to energetic sources will govern the chemical mechanisms at play in star-forming regions.  We have conducted millimeter imaging studies of Orion-KL with various beam sizes using CARMA in order to investigate the continuum structure. These $\lambda$=3mm observations have synthesized beam sizes of $\sim0.5\arcsec-5.0\arcsec$. These observations reveal the complex continuum structure of this region, which stands in sharp contrast to the previous structural models assumed for Orion-KL based on lower spatial resolution images.  The new results indicate that the spatial scaling previously used in determination of molecular abundances for this region are in need of complete revision.  Here we present the results of the continuum observations, discuss the sizes and structures of the detected sources, and suggest an observational strategy for determining the proper spatial scaling to accurately determine molecular abundances in the Orion-KL region.

\end{abstract}


\section{Introduction}
The Orion-KL region is the closest ($\sim$414 pc) site of massive star formation to Earth \citep{menten07}. There are several cloud components (e.g. Hot Core, Compact Ridge, extended ridge, and plateau) that are associated with Orion-KL, and these sources have varying chemical and physical properties \citep[e.g.][]{blake87}. The most chemically-rich components, including the Hot Core, IRc7, IRc6, and IRc5, are separated by less than $\sim$4000 AU (projected on the sky). Emission lines from large oxygen-bearing species (e.g.\ methyl formate [\mef] and dimethyl ether [\dme]) have been observed primarily toward IRc5 and IRc5 in the Compact Ridge, while emission from large nitrogen-bearing species (e.g.\ ethyl cyanide [\etcn]) is observed toward the Hot Core and IRc7 \citep{friedel08}.

The chemical pathways leading to the formation of complex organic molecules in regions such as Orion-KL are very poorly understood \citep[e.g][]{quan07}. It was once assumed that gas-phase ion-molecule reactions in hot cores drive most of the organic chemistry observed in interstellar clouds, but it has since been shown that such reactions are inefficient for producing many of the most highly-abundant molecules observed in these regions \citep{Garrod08}.  Instead, it is thought that energetic processing of interstellar ices is a likely driving force that dramatically influences interstellar chemistry \citep{Garrod08,Laas11}.  The spatial distribution of a given molecular species relative to the position of energetic sources (i.e. shock fronts, photodissociation regions, stellar objects, etc.) might offer clues to the processes that drive its formation. Previous interferometric observations of Orion-KL (see \citet{liu02} for an example) used older generation interferometers that provided insufficient spatial resolution to distinguish these individual regions and pinpoint possible correlations between a particular molecule and any given source.  \citet{friedel08} reported higher spatial resolution observations of several organic molecules toward Orion-KL at $\lambda$ = 1 mm. The results of these observations indicate a relationship between the type of physical environment and the chemistry observed in Orion-KL. Other recent observations targeting methyl formate at a similar spatial resolution reveal a highly complicated source structure \citep{favre11}. These recent results warrant further investigation at even higher spatial resolution.

To this end, we have carried out interferometric observations at the Combined Array for Research in Millimeter-Wave Astronomy (CARMA) over a range of spatial resolutions toward Orion-KL to more fully investigate the spatial extent of molecules in the region.  These observations targeted specific emission lines of several complex organic molecules.  Continuum observations at $\lambda$ = 3 mm  were conducted to explore source structure and size.  Below we overview these observations, present the continuum images at several beam sizes, and discuss the implications of these results for the standard spatial models assumed when determining molecular abundances for the Orion-KL region.

\section{Observations}
The observations were conducted in 2007 December, 2008 July, 2009 January, and 2010 April with the CARMA observatory in its B, D, A, and C configurations, respectively. These observations included two 7 hour tracks in B configuration, one 6 hour track in D configuration, four 5-6 hour tracks in A configuration, and three 4.5 hour tracks in C configuration. The observations have a phase center of $\alpha$(J2000) = $05^h35^m14^s.35$ and $\delta$(J2000) = $-05{\degr}22{\arcmin}35{\arcsec}.0$. The typical synthesized beams are $\sim5.9\arcsec\times4.8\arcsec$ (D configuration), $\sim2.2\times2.0\arcsec$ (C configuration), $\sim1.1\arcsec\times0.9\arcsec$ (B configuration), and $\sim0.4\times0.35\arcsec$ (A configuration). The $u-v$ coverage of the observations gives projected baselines of 3.0-35.3 k$\lambda$ (10-118 m, D configuration), 3.9-91.0 k$\lambda$ (13-304 m, C configuration), 21.8-257.5 k$\lambda$ (73-860 m, B configuration), and 38.0-467.2 k$\lambda$ (127-1560 m, A configuration). By observing over such a wide range of $u-v$ coverage, only structures larger than $\sim$30 arcseconds are resolved out. Each arcsecond is equivalent to $\sim$414 AU at the distance of Orion-KL.

The correlator was configured with six 31 MHz wide windows for continuum and spectral lines (three in each sideband) for the observations conducted in the B and D configurations. Each window had 63 channels with a channel spacing of 488 kHz ($\sim$1.4 \kms). The continuum images were constructed from both line-free windows and windows containing flagged lines. Uranus and Mars were used as flux density calibrators, and 0541-056 was used to calibrate the antenna-based gains. For two of the A configuration tracks, the correlator was configured with four 31 MHz wide windows and two 500 MHz wide windows (three in each sideband). The correlator configuration for the third A configuration track was two 31 MHz bands and four 500 MHz bands, while the fourth track was configured with four 31 MHz bands and two 62 MHz bands. The antenna-based gain calibration was done by self-calibrating on the SiO maser in source I at 86.243 GHz. The solution was then bootstrapped to the other bands. Phase offsets between each band and the SiO band were calculated and removed by using observations of 0607-085. As all available flux calibrators were heavily resolved in A configuration, the amplitude gains were calculated by comparing previously measured fluxes of 0607-085 with these observations. The absolute amplitude calibrations of 0541-056 and 0607-085 are accurate to within $\sim$20\%. The internal noise source was used to correct the passbands of each 31 MHz window, and observations of 0423-013 were used to correct the passbands of the 500 and 62 MHz windows. The data were calibrated, continuum subtracted, and imaged using the MIRIAD software package \citep{sault95}. Note that the flux of the point source BN (see $\S$\ref{sec:r-d}) is similar across all array configurations, indicating that the accuracy of the absolute flux calibration between the different tracks is better than 95\%.

\section{Results and Discussion}\label{sec:r-d}
\begin{figure}[!ht]
\includegraphics[scale=0.8]{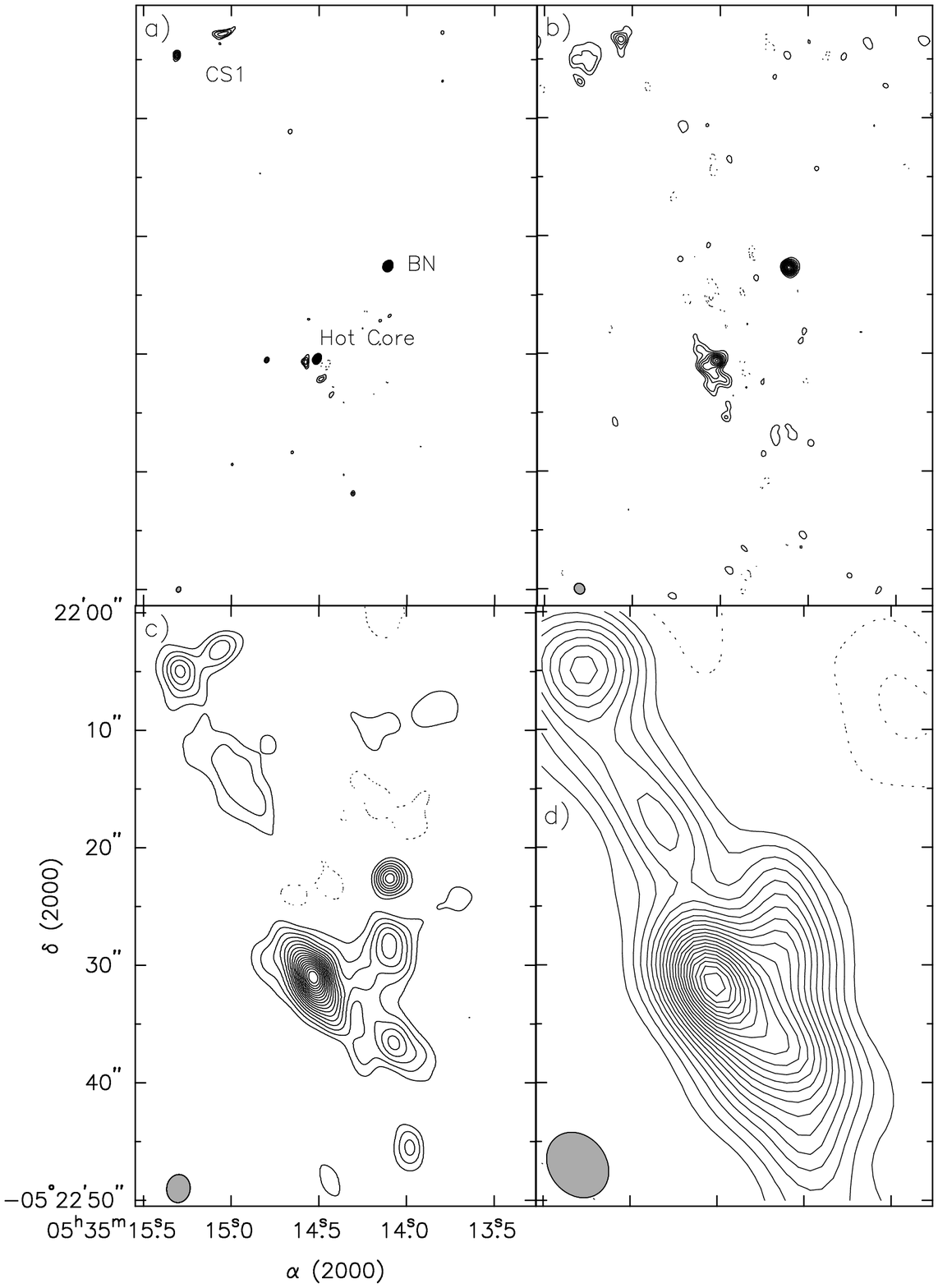}
\caption{Continuum maps of Orion-KL at $\lambda$ = 3 mm. a) CARMA A configuration map, with contours at $\pm$12$\sigma$, $\pm$18$\sigma$, $\pm$24$\sigma$, ... ($\sigma$ = 290 $\mu$\jbm); b) B configuration map, with contours at $\pm$6$\sigma$, $\pm$9$\sigma$, $\pm$12$\sigma$, ... ($\sigma$ = 2.4 m\jbm); c) C configuration map, with contours at $\pm9\sigma$, $\pm15\sigma$, $\pm21\sigma$, ... ($\sigma$ = 2.2 m\jbm); d) D configuration map, with contours at $\pm$12$\sigma$, $\pm$24$\sigma$, $\pm$36$\sigma$, ... ($\sigma$ = 1.8 m\jbm). Objects of note are labeled in the A configuration panel.  The synthesized beam size is shown in the lower left hand corner of each panel. \label{fig:contin}}
\end{figure}

Figure~\ref{fig:contin} shows the $\lambda$ = 3 mm continuum of Orion-KL from the A, B, C, and D array configuration observations. The noise levels are 290 $\mu$\jbm, 2.4 m\jbm, 2.2 m\jbm, and 1.8 m\jbm, respectively, and the synthesized beams for each array configuration are shown in the lower left corner of each panel. The large scale structures seen in the 5$\arcsec$ resolution observations are resolved into several dozen sources as the beam area decreases by two orders of magnitude. The strongest emission peak in the A and B configuration maps arises from source BN, whereas the peak of the emission in the C and D configuration maps is closely associated with the Hot Core. This indicates that, while there are numerous individual sources, the bulk of the continuum emission is coming from more extended regions.

\begin{figure}[!ht]
\includegraphics[angle=270,scale=0.95]{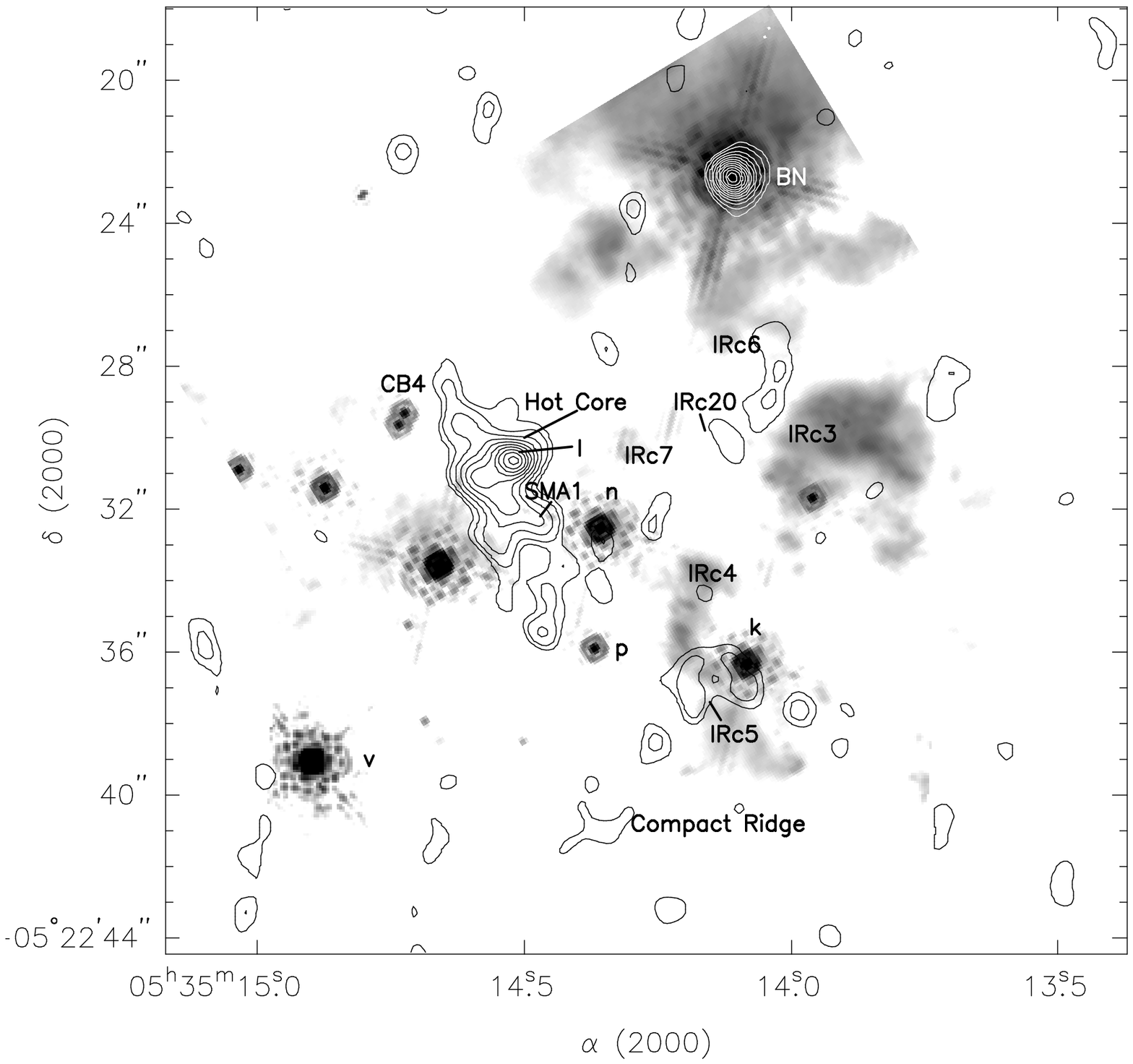}
\caption{ B configuration continuum image at $\lambda$=3 mm for the Orion-KL region overlaid on a 2 $\mu$m Hubble NICMOS image. Contours are 4$\sigma$, 6$\sigma$, 9$\sigma$, 12$\sigma$, ... ($\sigma$ = 2.4 m\jbm). Sources of interest have been labeled. Stars are labeled next to their position and all other sources are noted with labels on their central position unless indicated with a line.\label{fig:h-contin}}
\end{figure}

Figure~\ref{fig:h-contin} shows a portion of the $\lambda$=3 mm B configuration continuum (contours) overlaid on a gray scale Hubble NICMOS 2 $\mu$m image\footnote{Based on observations made with the NASA/ESA Hubble Space Telescope, obtained from the data archive at the Space Telescope Science Institute. STScI is operated by the Association of Universities for Research in Astronomy, Inc. under NASA contract NAS 5-26555.} of the same region. Sources of interest have been labeled for reference. Most of the millimeter continuum emission does not coincide with infrared active regions, but appears to be anti-correlated with most of the sources; the primary exception to this trend is source BN. Here we provide brief descriptions of previously identified infrared and millimeter sources of interest:

\begin{description}
\item[BN] A star with the highest $\lambda$ = 3 mm point source flux in the region (89 m\jbm).  This source is also very bright in the infrared (see Figure~\ref{fig:h-contin}).
\item[Hot Core/I] A site of high-mass star formation that is the traditional location of N-bearing molecules. Source I ($\sim$0.5\arcsec\ south of the hot core) is located at the center of the SiO masers \citep{menten95}.
\item[Compact Ridge] A region of stellar outflow interacting with the ambient cloud.  This source is the traditional location of O-bearing molecules in the region.   \cite{friedel08} noted that most of the compact emission comes from other sources. No continuum emission above the 3-$\sigma$ level was detected from this region in any of our observations.
\item[SMA1] A high-mass protostellar source \citep{beuth06}.
\item[n] A Herbig Ae/Be or mid B star with a luminosity $\sim2000L_\sun$ \citep{green04}. There is also a NH$_3$ peak, indicating a high ($>10^7$ cm$^{-3}$) density \citep{min89}. Weak emission ($\sim$10 m\jbm\ ) is seen toward this source in the B configuration results.
\item[p] A star which is detected both in the infrared and optical \citep{simp06}, with no detected $\lambda$ = 3 mm continuum.
\item[k] A star which is detected both in the infrared and optical \citep{simp06}, with no detected $\lambda$ = 3 mm continuum.
\item[v] A star which is detected both in the infrared and optical \citep{simp06}, with no detected $\lambda$ = 3 mm continuum.
\item[IRc3, 4 \& 5] These are objects thought to be reflection nebulae being illuminated by IRc2 (nearly coincident with source I) \citep{simp06}. There is no significant NH$_3$ emission seen toward any of these sources \citep{min89}. There is no detected continuum emission above 3 $\sigma$ toward IRc3, while a weak $\sim$9 m\jbm\ peak is detected toward IRc4. An apparent double-lobed continuum emission peak is centered in close proximity to IRc5.
\item[IRc6] There is notable $\lambda$=3 mm continuum emission coming from the region between IRc3 and IRc6. \citet{friedel08} also detected notable $\lambda$ = 1 mm continuum from this region.
\item[IRc7] This source is thought to contain an embedded YSO \citep{simp06} and has a strong NH$_3$ peak, indicating high density \citep{min89}.  However, there is no detected continuum emission above 3 $\sigma$. \citet{friedel08} reported weak $\lambda$ = 1 mm continuum from this source.
\item[IRc20] Little is known of this object because of its weak emission and proximity to the very bright BN emission.
\item[CB4] A binary with no detected $\lambda$ = 3 mm continuum.
\end{description}

\clearpage
\begin{figure}[t!]
\includegraphics[scale=0.8]{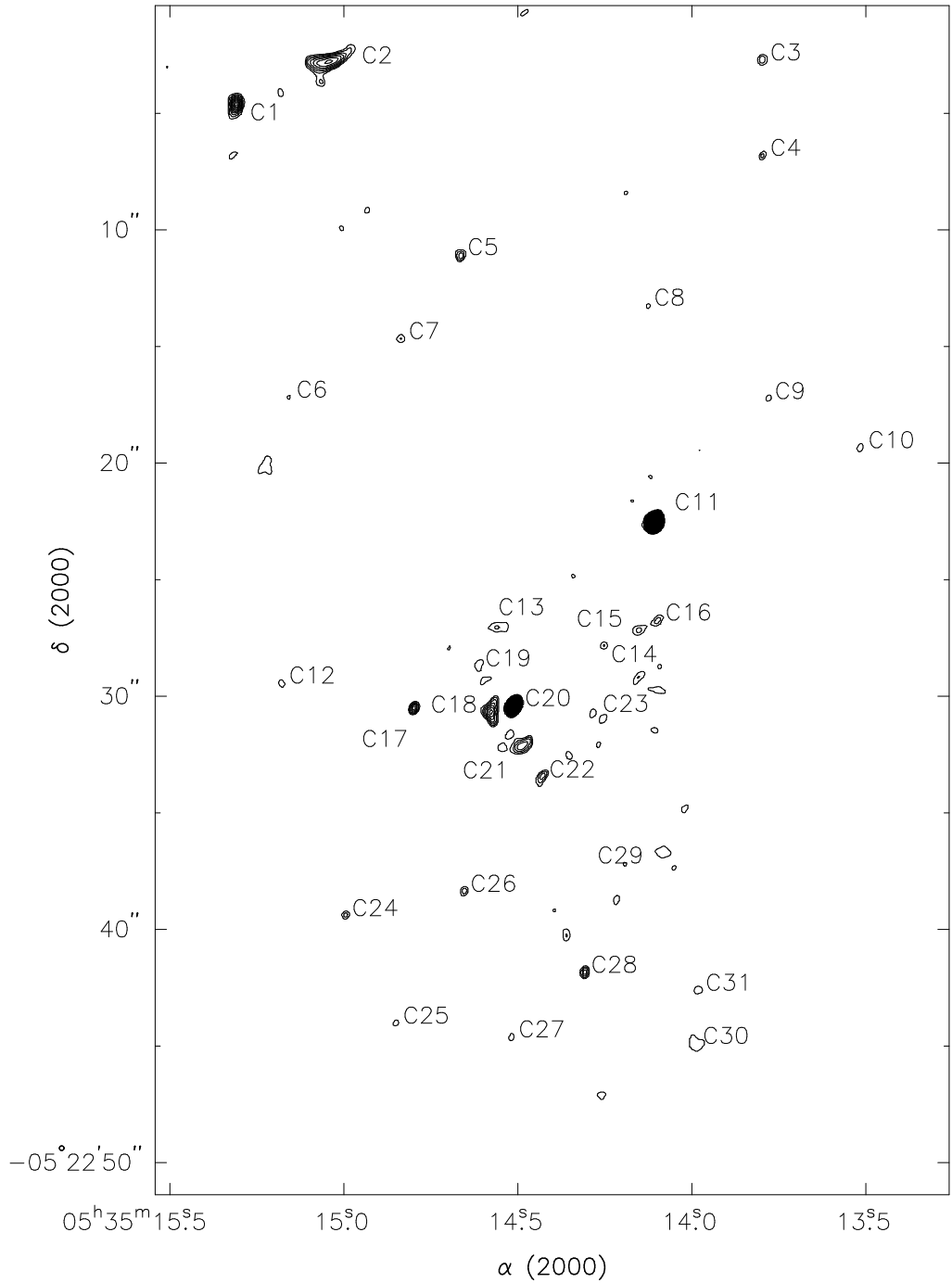}
\caption{The A configuration continuum of the region with sources labeled according to the designations in Table~\ref{tab:names}. The contours are 9$\sigma$, 12$\sigma$, 15$\sigma$,... ($\sigma$ = 280 $\mu$\jbm). \label{fig:names}}
\end{figure}
\clearpage

\begin{figure}[t!]
\includegraphics[scale=0.90]{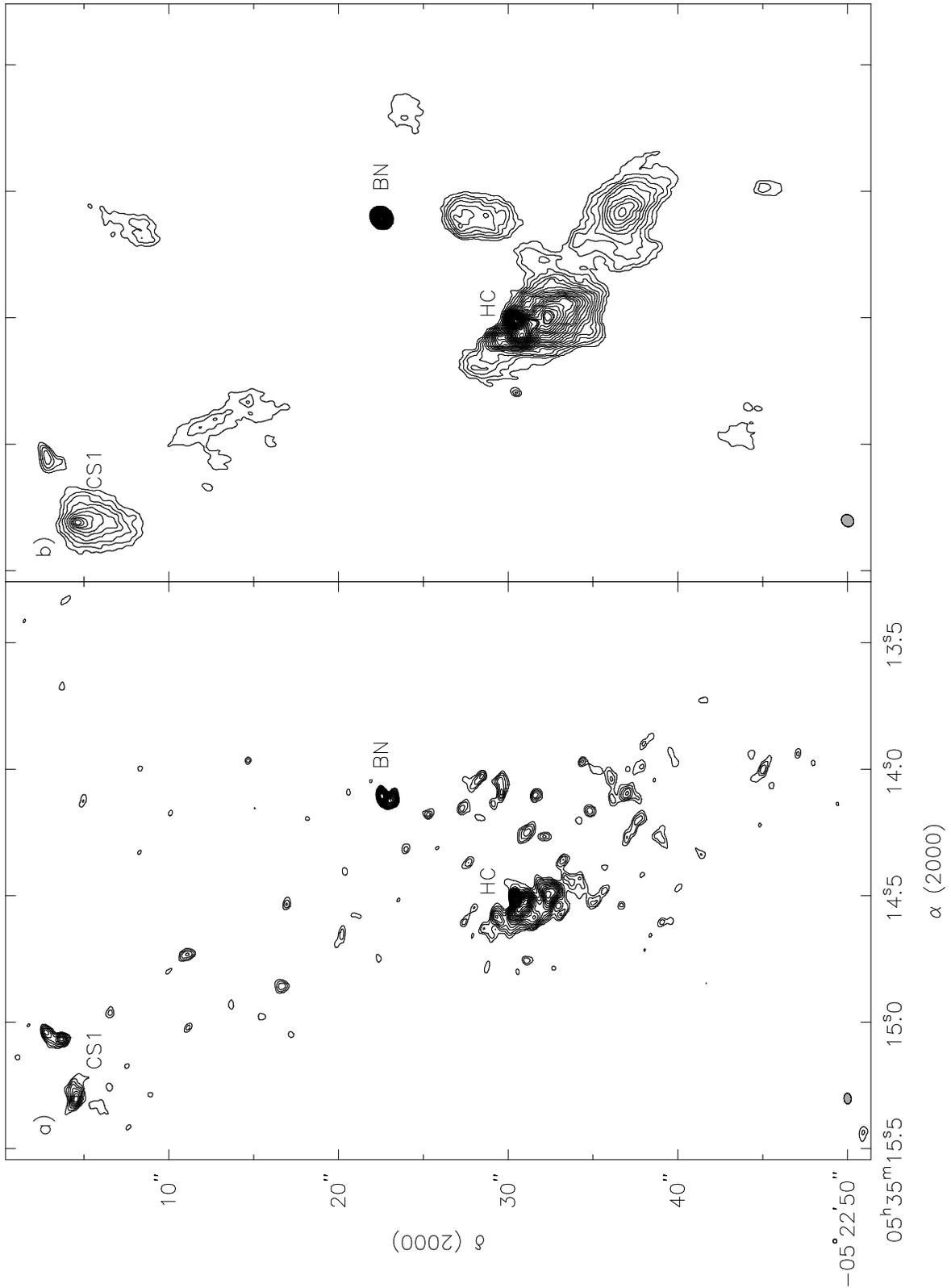}
\caption{a) Uniformly weighted map of all data. b) Naturally weighted map of all data. Sources of interest are labeled (HC is Hot Core)\label{fig:combined}}
\end{figure}
\clearpage

Figure~\ref{fig:names} shows the A configuration continuum with sources labeled. This is the most sensitive $\lambda$ = 3 mm map of this region to date, with a noise level of 290 $\mu$\jbm. The coordinates and source properties from the labeled sources in the map are given in Table~\ref{tab:names}. The first column gives the corresponding source ID from the map; the second and third columns give the coordinates; the fourth, fifth, and sixth columns give the deconvolved source size, position angle, and peak flux; and the seventh column gives the results of searching the SIMBAD online database. Figures~\ref{fig:names}b and \ref{fig:names}c show contour maps with all A, B, C, and D configuration data combined. The data for Figure~\ref{fig:combined}a were mapped using uniform weighting, while those for Figure~\ref{fig:combined}b were mapped using natural weighting. The synthesized beams for each map are shown in the lower left corner of each panel.

\begin{deluxetable}{lllcrrl}
 \rotate
\tablecolumns{7}
\tabletypesize{\scriptsize}
\tablewidth{0pt}
\tablecaption{Continuum Sources From the A Configuration}
\tablehead{\colhead{} & \colhead{} & \colhead{} & \colhead{Size\tablenotemark{b}} & \colhead{PA\tablenotemark{b}} & \colhead{Peak Flux\tablenotemark{b}} & \colhead{} \\
\colhead{ID} & \colhead{$\alpha$(J2000)\tablenotemark{a}} & \colhead{$\delta$(J2000)} & \colhead{($\arcsec\times\arcsec$)} & \colhead{($\degr$)} & \colhead{(m\jbm)} & \colhead{SIMBAD Results\tablenotemark{c}}}
\startdata
C1  & $05^h35^m15^s.310$  & $-05{\degr}22{\arcmin}04{\arcsec}.680$  & 0.72$\times$ 0.33 & 2.3 & 10.83 (93) & MM6 \citep{eisner06}\\
C2  & $05^h35^m15^s.047$  & $-05{\degr}22{\arcmin}02{\arcsec}.840$  & 2.1$\times$ 0.41 & -70.4 & 8.7 (82) & MM5 \citep{eisner06}\\
C3  & $05^h35^m13^s.796$  & $-05{\degr}22{\arcmin}02{\arcsec}.711$  & 0.33$\times$ 0.3 & 28.7 & 3.98 (32) & YSO? (138-203) \citep{odell96}\\
C4  & $05^h35^m13^s.796$  & $-05{\degr}22{\arcmin}06{\arcsec}.807$  & PS\tablenotemark{c} &  & 3.97 (31) & YSO (138-207) \citep{simp06}\\
C5  & $05^h35^m14^s.665$  & $-05{\degr}22{\arcmin}11{\arcsec}.061$  & 0.4$\times$ 0.29 & 41.6 & 4.73 (76) & YSO (147-211) \citep{doi02}\\
C6  & $05^h35^m15^s.156$  & $-05{\degr}22{\arcmin}17{\arcsec}.173$  & 0.48$\times$ 0.18 & -60.0 & 2.52 (53) & H$_2$O maser \citep{gaume98}\\
C7  & $05^h35^m14^s.835$  & $-05{\degr}22{\arcmin}14{\arcsec}.660$  & 0.42$\times$ 0.19 & 46.5 & 3.37 (51) & No sources within 2$\arcsec$ radius\\
C8  & $05^h35^m14^s.129$  & $-05{\degr}22{\arcmin}13{\arcsec}.228$  & 0.64$\times$ 0.18 & 74.1 & 2.84 (40) &  Star? (9) \citep{simp06}\\
C9  & $05^h35^m13^s.782$  & $-05{\degr}22{\arcmin}17{\arcsec}.235$  & 0.66$\times$ 0.48 & -26.8 & 2.79 (49) &  Star? COUP J053513.7-052217 \citep{simp06}\\
C10 & $05^h35^m13^s.515$  & $-05{\degr}22{\arcmin}19{\arcsec}.340$  & 0.30$\times$ 0.15 & -12.3 & 3.42 (23) & YSO (135-220) \citep{simp06}\\
C11 & $05^h35^m14^s.106$  & $-05{\degr}22{\arcmin}22{\arcsec}.528$  & 0.11$\times$ 0.02 & 7.3 & 90.76 (89) & Embedded star (BN) \citep{BN}\\
C12 & $05^h35^m15^s.179$  & $-05{\degr}22{\arcmin}29{\arcsec}.431$  & 0.40$\times$ 0.06 & 43.8 & 3.15 (26) &  X-ray source? \citep{feig02}\\
C13 & $05^h35^m14^s.556$  & $-05{\degr}22{\arcmin}27{\arcsec}.128$  & 0.97$\times$ 0.36 & -87.8 & 4.89 (29) & No sources within 2$\arcsec$ radius\\
C14 & $05^h35^m14^s.253$  & $-05{\degr}22{\arcmin}27{\arcsec}.820$  & PS\tablenotemark{d} &  & 5.27 (44) & IRc6E \citep{shu04}\\
C15 & $05^h35^m14^s.155$  & $-05{\degr}22{\arcmin}27{\arcsec}.193$  & 0.85$\times$ 0.38 & -47.7 & 4.9 (20) & IRc6 \citep{rieke73}\\
C16 & $05^h35^m14^s.096$  & $-05{\degr}22{\arcmin}26{\arcsec}.734$  & PS\tablenotemark{d} &  & 4.43 (29) & IRc6N \citep{gezari98}\\
C17 & $05^h35^m14^s.797$  & $-05{\degr}22{\arcmin}30{\arcsec}.557$  & PS\tablenotemark{d} &  & 9.23 (57) & H$_2$O maser \citep{gaume98}\\
C18 & $05^h35^m14^s.579$  & $-05{\degr}22{\arcmin}30{\arcsec}.679$  & 0.61$\times$ 0.28 & 10.8 & 7.08 (94) & YSO (146-231)/H$_2$O maser \citep{simp06,gaume98}\\
C19 & $05^h35^m14^s.614$  & $-05{\degr}22{\arcmin}28{\arcsec}.696$  & 0.59$\times$ 0.31 & -18.2 & 4.39 (43) & MM24 \citep{eisner08}\\
C20 & $05^h35^m14^s.511$  & $-05{\degr}22{\arcmin}30{\arcsec}.397$  & PS\tablenotemark{d} &  & 50.09 (97) &  Source I \citep{church87}\\
C21 & $05^h35^m14^s.488$  & $-05{\degr}22{\arcmin}32{\arcsec}.101$  & 0.87$\times$ 0.44 & -64.7 & 6.05 (60) &  SMA1 \citep{beuth06}\\
C22 & $05^h35^m14^s.431$  & $-05{\degr}22{\arcmin}33{\arcsec}.457$  & 0.70$\times$ 0.17 & -48.1 & 5.31 (75) & Hot Core\\
C23 & $05^h35^m14^s.278$  & $-05{\degr}22{\arcmin}30{\arcsec}.647$  & 0.47$\times$ 0.31 & -23.4 & 4.18 (9) & IRc7 \citep{ww84}\\
C24 & $05^h35^m14^s.996$  & $-05{\degr}22{\arcmin}39{\arcsec}.411$  & 0.29$\times$ 0.19 & -87.7 & 5.39 (34) & YSO (150-240) \citep{lada04}\\
C25 & $05^h35^m14^s.850$  & $-05{\degr}22{\arcmin}44{\arcsec}.020$  & 0.35$\times$ 0.27 & -70.3 & 2.98 (10) & IR source \citep{luhman00}\\
C26 & $05^h35^m14^s.656$  & $-05{\degr}22{\arcmin}38{\arcsec}.363$  & PS\tablenotemark{d} &  & 5.4 (45) & Star?/YSO (149-239) \citep{lada04}\\
C27 & $05^h35^m14^s.519$  & $-05{\degr}22{\arcmin}44{\arcsec}.609$  & 0.68$\times$ 0.30 & -20.9 & 2.87 (22) & IR source (MLLA 554) \citep{muench02}\\
C28 & $05^h35^m14^s.309$  & $-05{\degr}22{\arcmin}41{\arcsec}.839$  & \tablenotemark{e} &  & 6.35 (56) & H$_2$O maser nearby \citep{gaume98}\\
C29 & $05^h35^m14^s.20$   & $-05{\degr}22{\arcmin}37{\arcsec}.15$   & \tablenotemark{f} &  &  & H$_2$O and OH maser \citep{braz83}\\
C30 & $05^h35^m13^s.988$  & $-05{\degr}22{\arcmin}44{\arcsec}.887$  & 0.91$\times$ 0.70 & 53.3 & 3.46 (33) & MM23 \citep{eisner06}\\
C31 & $05^h35^m13^s.981$  & $-05{\degr}22{\arcmin}42{\arcsec}.611$  & 0.58$\times$ 0.38 & 78.2 & 3.18 (15) & CH$_3$OH maser \citep{johnston92}\\
\enddata

\tablenotetext{a}{Peak coordinates have an uncertainty $\leq$ the $\sim$0.4$\arcsec$ synthesized beam.}
\tablenotetext{b}{Determined the from the MIRIAD task IMFIT.  Source sizes and position angle given here are the deconvolved values.}
\tablenotetext{c}{Obtained from a coordinate search in the SIMBAD database (http://simbad.harvard.edu/simbad/).}
\tablenotetext{d}{The deconvolution resulted in a point source.}
\tablenotetext{e}{The deconvolution failed, thus no source size is reported.}
\tablenotetext{f}{A reliable fit could not be obtained; the source position is estimated from its peak.}

\label{tab:names}
\end{deluxetable}

Table~\ref{tab:flux} shows a comparison of the integrated fluxes for the detected sources between the different array configurations. The first column gives the source ID; the second column gives the integrated flux from the D configuration data; the third and fourth columns give the integrated flux and fitted size from the C configuration data; the fifth and sixth columns give the percentage of resolved flux and fitted source size from the B configuration data; and the seventh column gives the percentage of resolved flux from the A configuration data. The percentage of resolved flux listed in both cases is in comparison to the C configuration data, as that is the largest beam size with which most of the sources can be resolved. The percentage of resolved flux was determined by convolving the B and A configuration data with the synthesized beam from the C configuration data (using the MIRIAD task CONVOL), and comparing the fitted integrated fluxes. Here, 0\% resolved out means no missing flux, while 100\% means all flux was resolved out. The last row of Table~\ref{tab:flux} is a combination of all continuum sources C18 - C22 as they are unresolved in both the C and D configuration data.

\begin{deluxetable}{l|lrccccc}
\tablecolumns{8}
 \rotate
\tabletypesize{\scriptsize}
\tablewidth{0pt}
\tablecaption{Comparison of Fluxes of Continuum Sources}
\tablehead{             \colhead{}   & \colhead{D Configuration\tablenotemark{a}} & \multicolumn{2}{c}{C Configuration\tablenotemark{b}}                                      & \colhead{} & \multicolumn{2}{c}{B Configuration\tablenotemark{c}}             & \colhead{A Configuration\tablenotemark{d}}\\
\cline{3-4} \cline{6-7} \colhead{}   & \colhead{$\int$Flux}               & \colhead{$\int$Flux}                         & \colhead{Size\tablenotemark{e}}    & \colhead{} & \colhead{Resolved out} & \colhead{Size\tablenotemark{e}} & \colhead{Resolved out}\\
                        \colhead{ID} & \colhead{(m\jbm)}                  & \colhead{(m\jbm)}                            & \colhead{($\arcsec\times\arcsec$)} & \colhead{} & \colhead{(\%)\tablenotemark{f}} & \colhead{($\arcsec\times\arcsec$)} & \colhead{(\%)\tablenotemark{f}}}
\startdata
C1   & \multirow{2}{*}{{$\left.\begin{matrix}~\\~\end{matrix}\right\}$}428.9 (14.7)\tablenotemark{g}} & 152.9 (6.7) & $2.5\times1.8$ && 0 & $3.2\times2.7$ & 49 \\
C2   &  & 98.8 (4.5) & $2.9\times0.3$ && 0 & $1.5\times0.87$ & 17 \\
C3   & \nodata & \nodata & \nodata && \nodata & \nodata & \nodata \\
C4   & \nodata & 145.7(10.7) & $7.2\times2.8$ && \nodata & \nodata & 75 \\
C5   & \nodata & \nodata & \nodata && \nodata & \nodata & \nodata \\
C6   & \nodata & \nodata & \nodata && \nodata & \nodata & \nodata \\
C7   & 548.0 (22.4) & 540.0(26.0) & $12.6\times4.4$ && \nodata & \nodata & 93 \\
C8   & \nodata & \nodata & \nodata && \nodata & \nodata & \nodata \\
C9   & \nodata & \nodata & \nodata && \nodata & \nodata & \nodata \\
C10  & \nodata & 14.1 (0.9) & $2.2\times7.6$ && \nodata & \nodata & 0 \\
C11  & 100.0\tablenotemark{h} & 107.4 (2.1) & PS\tablenotemark{i} && 0 & $0.51\times0.34$ & 0 \\
C12  & \nodata & \nodata & \nodata && \nodata & \nodata & \nodata \\
C13  & \nodata & \nodata & \nodata && \nodata & \nodata & \nodata \\
C14  & \multirow{3}{*}{$\left.\begin{matrix}~\\~\\~\end{matrix}\right\}$611.8 (100.9)\tablenotemark{j}} & \nodata & \nodata && \nodata & \nodata & \nodata \\
C15  &  & 129.3 (92.2) & $5.2\times1.1$\tablenotemark{k} && \nodata & \nodata & 40 \\
C16  &  & 157.2 (29.9) & $2.5\times2.2$\tablenotemark{k} && \nodata & \nodata & \nodata \\
C17  & \multirow{6}{*}{$\left.\begin{matrix}~\\~\\~\\~\\~\\~\end{matrix}\right\}$913.1\tablenotemark{l}} & 137.0 (12.9) & $3.5\times2.9$ && \nodata & \nodata & \nodata \\
C18  &  & 278.9\tablenotemark{k} & $5.0\times4.2$ && 47 & $1.4\times1.2$ & \nodata \\
C19  &  & 146.7\tablenotemark{k} & $4.0\times2.9$ && 88 & \nodata & \nodata \\
C20  &  & 122.9\tablenotemark{k} & PS\tablenotemark{i} && 43 & $0.69\times0.39$ & \nodata \\
C21  &  & 258.2\tablenotemark{k} & $5.1\times3.7$ && 51 & $0.80\times0.66$ & \nodata \\
C22  &  & 135.0\tablenotemark{k} & $3.3\times3.0$ && 19 & $1.3\times0.57$ & \nodata \\
C23  & \nodata & 223.9 (5.8) & $4.4\times3.4$ && \nodata & \nodata & 71 \\
C24  & \nodata & 14.4 (1.0) & PS\tablenotemark{i} && 0 & PS\tablenotemark{i} & \nodata \\
C25  & \nodata & \nodata & \nodata && \nodata & \nodata & \nodata \\
C26  & \nodata & \nodata & \nodata && \nodata & \nodata & \nodata \\
C27  & \nodata & \nodata & \nodata && \nodata & \nodata & \nodata \\
C28  & \nodata & 17.2 (1.7) & PS\tablenotemark{i} && \nodata & \nodata & \nodata \\
\multirow{2}{*}{C29}  & \multirow{2}{*}{551.9 (181.9)\tablenotemark{m}} & \multirow{2}{*}{112.5 (5.2)} & \multirow{2}{*}{$3.2\times2.0$} && \multirow{2}{*}{0\tablenotemark{n}{$~\left\{\begin{matrix}~\\~\end{matrix}\right.$}} & $2.5\times0.53$ & \multirow{2}{*}{7} \\
     &  &  &  &&  & $2.0\times0.8$ & \\
C30  & \nodata & 80.4 (4.1) & $2.5\times0.8$ && 0 & \nodata & 30 \\
C31  & \nodata & \nodata & \nodata && \nodata & \nodata & \nodata \\
C32\tablenotemark{o}  & \nodata & 238.2 (8.1) & $4.3\times2.2$ && \nodata & \nodata & 65 \\
C33\tablenotemark{p}  & \nodata & \nodata & \nodata && \nodata & $1.2\times0.83$ & \nodata \\
C34\tablenotemark{q}  & \nodata & \nodata & \nodata && \nodata & $0.80\times0.30$ & \nodata \\
C18-C22\tablenotemark{r} & 913.2 (5.4) & 829.3 (40.5) & \nodata && 46 & \nodata & 82 \\
\enddata

\tablenotetext{a}{Synthesized beam is $5.9\arcsec\times4.9\arcsec$}
\tablenotetext{b}{Synthesized beam is $2.4\arcsec\times2.0\arcsec$.}
\tablenotetext{c}{Synthesized beam is $0.96\arcsec\times0.85\arcsec$}
\tablenotetext{d}{Synthesized beam is $0.47\arcsec\times0.37\arcsec$}
\tablenotetext{e}{Size was determined from the MIRIAD task IMFIT. Reported values are the deconvolved sizes.}
\tablenotetext{f}{Percentage of the source flux that was resolved out in comparison to C configuration observations, 0\% means no flux resolved out, 100\% means all flux resolved out.}
\tablenotetext{g}{At this resolution, C1 \& C2 are completely unresolved, the reported integrated flux is from a single component fit.}
\tablenotetext{h}{A reliable fit could not be obtained due to the large synthesized beam. The integrated flux was estimated from the peak intensity and expected point source size.}
\tablenotetext{i}{The deconvolution resulted in a point source.}
\tablenotetext{j}{The integrated flux is a summation of C14-C16, as the individual sources could not be resolved.}
\tablenotetext{k}{Fit results were poor and no uncertainty was given.}
\tablenotetext{l}{The integrated flux is a summation of C17-C22, as the individual sources could not be resolved.}
\tablenotetext{m}{The integrated flux may include contributions from C23, as the individual sources could not be resolved.}
\tablenotetext{n}{In the B configuration two sources are associated with C29 (see Figure~\ref{fig:contin}), the resolved flux percentage is a combination of both sources.}
\tablenotetext{o}{This continuum source was only detected in the C configuration, peak position is $05^h35^m14^s.070$, $-05{\degr}22{\arcmin}36{\arcsec}.581$.}
\tablenotetext{p}{This continuum source was only detected in the B configuration, peak position is $05^h35^m14^s.613$, $-05{\degr}22{\arcmin}29{\arcsec}.688$.}
\tablenotetext{q}{This continuum source was only detected in the B configuration, peak position is $05^h35^m14^s.464$, $-05{\degr}22{\arcmin}35{\arcsec}.350$.}
\tablenotetext{r}{The values reported here are for a sum total from all continuum sources C18-C22.}

\label{tab:flux}
\end{deluxetable}

\FloatBarrier

The following is a discussion of the individual continuum sources detected. Any source not listed below was only detected by the A configuration observations, indicating it is a lone point-like source, or that it has a surrounding envelope below our detection threshold.
\begin{description}
\item[C1 \& C2] Both of these sources appear to be embedded in an extended envelope, on the order of $\sim10\arcsec$, with the bulk of the flux surrounding C1. A third of the flux is resolved out by a 2\arcsec\ beam, and a total of 63\% is resolved out by the subarcsecond beam.
\item[C4] This continuum source appears to be a single point source embedded in a surrounding envelope several arcseconds in diameter, which emits 75\% of the flux. This source was undetected in the D configuration observations due to its proximity to the large negative sidelobes of the main continuum sources.
\item[C7] This source appears to bridge the gap between C1/C2 and the Hot Core region, and little flux is lost with a 2\arcsec\ beam. However, 93\% of the flux is lost with the subarcsecond A configuration beam, indicating that the small C7 source is surrounded by a large$\sim12\arcsec\times4\arcsec$ envelope.
\item[C10] This source is detected with the $\sim$2\arcsec\ C configuration beam with a source size of $\sim2\arcsec\times7\arcsec$, however there is no flux lost when observing with a subarcsecond beam, indicating that the near point source may be embedded in a very diffuse envelope.
\item[C11] This source is BN and is expected to be a point source, as can be seen by the fact that there is no flux lost from the lowest to highest resolution observations.
\item[C13 \& C14] These sources could not be individually resolved from surrounding sources in all but the A configuration, and so no additional properties can be inferred.
\item[C15 \& C16] Both of these sources appear to be embedded in an extended envelope, on the order of $\sim5\arcsec$. Over half of the flux is resolved out by a 2\arcsec\ beam, and a total of $\sim$80\% is resolved out by the subarcsecond beam.
\item[C17 - C22] The individual sources could not be fully resolved by all but the highest resolution observations. Table~\ref{tab:flux} gives the fitted parameters for the B and C configurations, however the errors are large due to source blending. The total flux of the complex (given in the last line of Table~\ref{tab:flux}) shows that there is a notable large envelope surrounding all of the sources, over half of which is resolved out when observed with a $\sim$2\arcsec\ beam, and 82\% of which is resolved out by the subarcsecond A configuration beam.
\item[C23] This source, like many others, appears to be a near point source embedded in a surrounding envelope of a few arcseconds in size which produces nearly three quarters of the continuum flux.
\item[C24] This source appears to be a point source with no surrounding envelope in all array configurations.
\item[C28] The fit for this source from the A configuration data is poor, but based on the map flux it appears to be a point source with a minimal surrounding envelope.
\item[C29] This source has notable extended flux, 75\% of which is resolved out by a $\sim$2\arcsec\ beam. In the B configuration data this source appears as two distinct sources on either side of IRc5, only one of which is detected by the subarcsecond beam.
\item[C30] This source was detected with all four array configurations but could not be resolved for fitting in the D configuration data.  It appears to be a small source embedded in a weakly emitting surrounding envelope.
\item[C32] This source was only detected in the C configuration and in the convolved A configuration data, indicating that it is a small source with a surrounding envelope that is below the detection threshold of the D and B configuration observations.
\item[C33 \& C34] These sources were only detected in the B configuration and appear to be small, but larger than point-like sources.
\end{description}

\begin{figure}[t!]
\includegraphics[angle=270,scale=0.8]{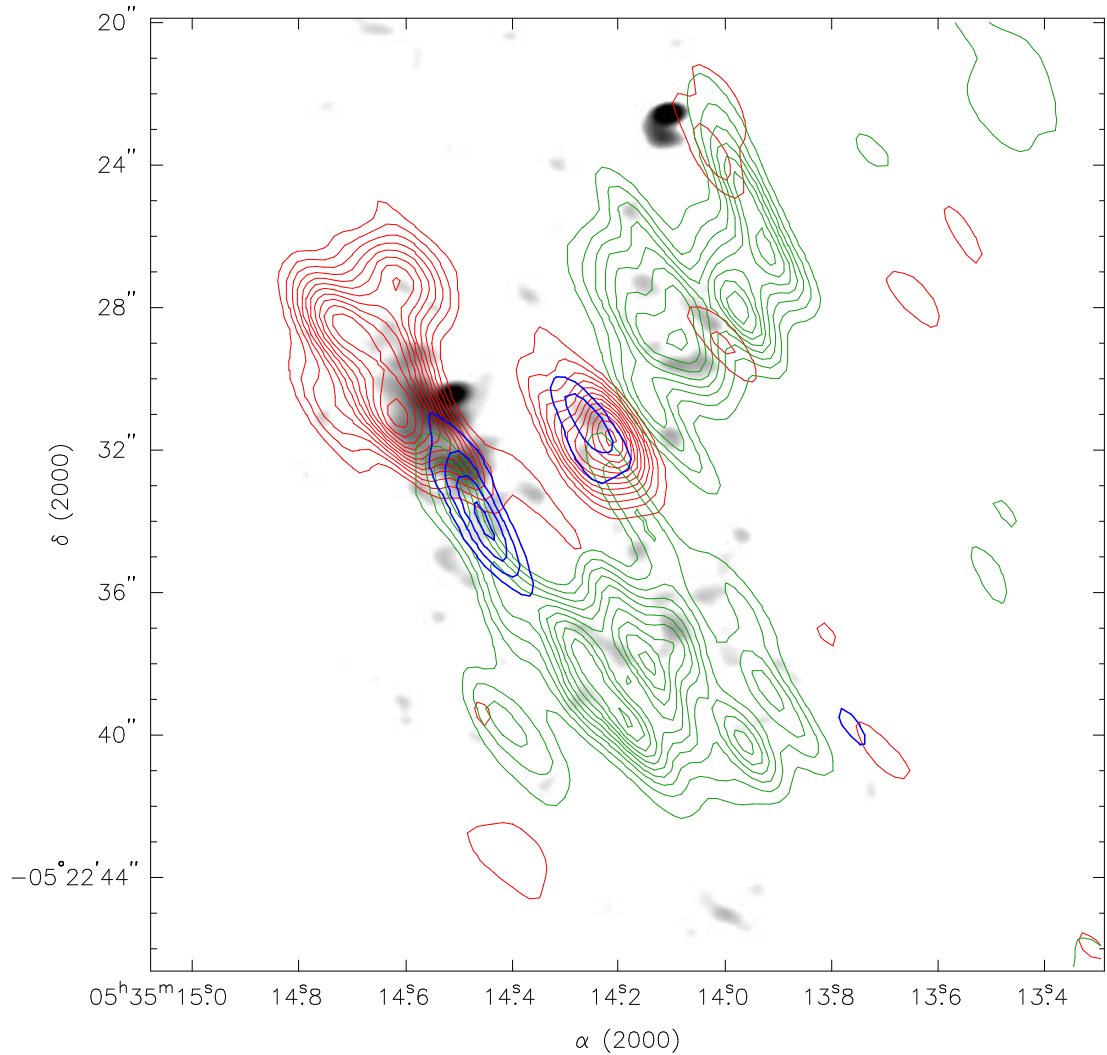}
\caption{Molecular emission contours from \citep{friedel08} overlaid on the gray scale continuum (uniformly weighted map). The continuum has been logarithmically scaled to show the weaker features in this high dynamic range image. The green contours are ethyl cyanide [\etcn], red contours are dimethyl ether [\dme], and the blue contours are acetone [\ace].\label{fig:overlay}}
\end{figure}

A standard rule-of-thumb with molecular line observations in Orion-KL, based on the images presented by \cite{liu02}, has been to assume a $\sim$5\arcsec $\times$ 5\arcsec ~spherical source structure for the Orion Hot Core, and an elongated $\sim$5\arcsec $\times$ 10\arcsec ~oval source structure for the Compact Ridge.  The results from the present work call into question this standard spatial model used to interpret Orion-KL molecular observations.  These observations reveal that the Orion Hot Core and Compact Ridge are comprised of many bright but compact continuum sources that have not been spatially resolved in previous interferometric studies.  These point sources are interspersed with regions of extended emission that are directly correlated with sources of heating and/or shocks.  Given the complexity of the continuum morphology in this region, it is likely that the spatial distribution of molecular line emission will be similarly complex. Further evidence of this complex source morphology is shown in Figure~\ref{fig:overlay}, which compares the continuum images derived from this work to the molecular line images of \cite{friedel08}. The gray scale traces the uniformly weighted continuum (logarithmically scaled to show the weaker features in this high dynamic range image), the green contours trace ethyl cyanide [\etcn], the red contours trace dimethyl ether [\dme], and the blue contours trace acetone [\ace].  This comparison reveals that significant source structure remains unresolved even with the 1\arcsec ~beam used in the \cite{friedel08} observations.  This comparison also reveals that the continuum emission and molecular emission do not always necessarily trace the same source morphology.  We conclude from this comparison that only direct, high spatial resolution observations for each individual molecule can be used to determine the true source morphology for that particular molecule.  This information can then be used to accurately determine molecular abundances in the Orion-KL region.

\section{Summary}
We have conducted extensive $\lambda$ = 3 mm continuum observations of the Orion-KL star forming region. These observations used the CARMA Array in four different array configurations to fully sample this region at varying spatial resolutions ($\sim$0.5\arcsec\ - 5\arcsec\ beams). It is apparent from these observations that the spatial distribution of the continuum in the Orion-KL region is much more complicated than has previously been considered. In the past it has been routine to assume that there were two main sources of continuum in Orion-KL: the Hot Core/source I, and source BN. In light of the results from the continuum observations reported here, this simplistic view of the Orion-KL structure must be revised.  These observations show that these sources are comprised of nearly three dozen individual continuum sources, many of which appear to be point-like sources surrounded by extended envelopes. It would not be surprising for future higher spatial resolution observations to detect even more small continuum sources in the region. Comparison of these continuum maps to molecular maps from \citet{friedel08} reveal that high spatial resolution studies are required for each molecule detected in the Orion-KL region before a proper source structure model can be determined.  Such studies will allow for more accurate determination of molecular abundances in the Orion-KL region.

\acknowledgements
We would like to thank an anonymous referee for their helpful comments. This work was partially funded by NSF grant AST-0540459 and the University of Illinois. Support for CARMA construction was derived from the states of Illinois, California, and Maryland, the Gordon and Betty Moore Foundation, the Kenneth T. and Eileen L. Norris Foundation, the Associates of the California Institute of Technology, and the National Science Foundation. Ongoing CARMA development and operations are supported by the National Science Foundation under a cooperative agreement, and by the CARMA partner universities. This research has made use of the SIMBAD database, operated at CDS, Strasbourg, France. We would also like to thank Dr. Richard Plambeck for sharing observational data.

\end{document}